\documentclass[11pt]{article}
\usepackage{geometry}                % See geometry.pdf to learn the layout options. There are lots.
\usepackage{graphicx}
\usepackage{amssymb}
\usepackage{epstopdf}
\usepackage{graphicx}
\usepackage{amsmath}
\DeclareMathAlphabet{\mathpzc}{OT1}{pzc}{m}{it}
\usepackage{bm}
\usepackage{mathrsfs}
\usepackage{hyperref}
\usepackage[toc,page]{appendix}
\DeclareMathAlphabet{\mathpzc}{OT1}{pzc}{m}{it}

\newcommand{\pr}[1]{\frac{\partial}{\partial #1}}

\newcommand {\susy}{\mathfrak{susy}}

\newcommand {\Hom}{\mathrm{Hom}}

\newcommand {\A}{\mathcal{A}}

\newcommand {\tr}{\mathrm{tr}}

\newcommand{\g}{\mathfrak{g}}

\newcommand{\Sym}{\mathrm{Sym}}

\newcommand{\ity}{_{\infty}}

\newcommand{\rH}{\mathrm{H}}
\newcommand{\rHH}{\mathrm{HH}}
\newcommand{\rHC}{\mathrm{HC}}
\newcommand{\cO}{\mathcal{O}}

\begin{document}

\title{Generalized Chern-Simons action and maximally supersymmetric gauge theories}

%\date{}                                           % Activate to display a given date or no date
\author{M. V. Movshev\\Stony Brook University\\Stony Brook, NY 11794-3651, USA
\\ A. Schwarz\\ Department of Mathematics\\ 
University of 
California \\ Davis, CA 95616, USA \thanks{The work of both  authors was partially supported by NSF 
grants}}
\begin{abstract}\large
We study observables and deformations of generalized Chern-Simons action and  show how to apply these results to maximally supersymmetric gauge theories. We describe a construction of  large class of deformations based on some results  on  the cohomology of  super Lie algebras proved in the Appendix.
\end{abstract}
\maketitle

%\section{}
%\subsection{}

\section { Chern-Simons functional.}

One can construct Chern-Simons functional 
for every differential associative ${\bf Z}_2$-graded 
algebra ${\mathcal A}$ equipped with trace $\rm tr.$ By definition trace is a linear functional vanishing on supercommutators; we assume that the trace is closed  (vanishes on elements of the form $da$). This linear functional can be odd or even; we are mostly interested in the case when it is  it is odd. 

 We can define 
 invariant  inner product  in terms 
of trace: $<a,b>={\rm tr} ab$.  Let us assume that this inner product is non-degenerate.

For every $N$ we 
define the associative algebra ${\mathcal A}_N$ as tensor 
product ${\mathcal A} \otimes {\rm Mat}_N$ where 
${\rm Mat}_N$ stands for the matrix algebra. (In other words, ${\mathcal A}_N$ is an algebra of $N\times N$ matrices with entries from ${\mathcal A}_N$.) Using the trace in $\mathcal A$ and the conventional matrix  trace we define the trace and invariant inner product in  ${\mathcal A}_N$.

Chern-Simons functional on  $\Pi {\mathcal A}_N$ is defined
by the formula
$$ CS_N(A)={1\over 2} {\rm tr}AdA+{2\over 3}  {\rm 
tr}A^3={1\over 2} {\rm tr}AdA+{1\over 3} {\rm 
tr}A[A,A]$$
Here $\Pi$ stands for the parity reversal. Chern-Simons functional is even if the trace has  odd parity,  as we will assume in what follows.  We omit the index $N$ in the notation $CS_N.$ 

Notice that in this definition we need really only the super Lie algebra 
structure  and invariant inner product in
the  algebra ${\mathcal A}_N$:
$$CS(A)={1\over 2}<A,dA>+{1\over 3}<A,[A,A]>.$$
The functional $CS$ coincides with the standard Chern-Simons functional  in the case when 
${\mathcal A}$ is the algebra $\Omega (M)$ of differential 
forms on three-dimensional manifold $M$ equipped with a trace ${\rm 
tr}C=\int_M C$.

 An odd non-degenerate  inner product on  vector space induces such a product on its dual that can be regarded as an odd symplectic form. This form $\omega=dz^a\omega_{ab}dz^b$ can be used to define an odd Poisson bracket on functions:
 $\{F,H\}=\overset{\leftarrow}{\partial} _aF\omega ^{ab}\overset{\rightarrow}{\partial}_bH,  $
 and to assign a vector field $\partial_bF\omega ^{ab}$ to every function. (This vector field corresponds to the first order differential operator $\xi_F$ defined by the formula $\xi_F(H)=\{F,H\}.)$
 
 Applying this remark to ${\mathcal A}_N$ we see that  the odd vector field $Q$ corresponding  to the functional $CS$ can be written in the form
 \begin{equation}
\label{Q}
\delta _QA=dA+\frac{1}{2}[A,A].
\end{equation}
 This vector field obeys $[Q,Q]=0$. {\footnote {An odd vector field $Q$ obeying $[Q,Q]=0$ is called homological vector field, because the corresponding first order differential operator $\hat Q$ obeys $\hat Q^2=0$ and therefore can be considered as a differential acting on the space of functions.}}.The relation $[Q,Q]=0$ is equivalent to the BV master equation $\{CS,CS\}=0.$ Notice that that vector field $Q$ does depend  on super Lie algebra $L_N{\mathcal A}=Lie {\mathcal A}_N$ corresponding to the associative algebra ${\mathcal A}_N$, but it does not depend of inner product.
 
The construction of  the  functional $CS$ can be generalized 
to the case when ${\mathcal A}$ is an 
$A_{\infty}$-algebra equipped with invariant inner 
product. Recall that the structure of $ 
A_{\infty}$-algebra 
 ${\mathcal A}$ on a $\mathbb{Z}_2$-graded space is specified 
by means of a sequence $^{(k)}m$ of operations, satisfying some relations that will be described later.  The operation  $^{(k)}m$ has $k$ arguments, in a coordinate system it is specified 
by a tensor $^{(k)}m^a_{a_1,...,a_k}$ having one upper 
index and $k$ lower indices. Having an  inner product we 
can lower the upper index; invariance of  inner product 
means that the tensor  $^{(k)}\mu _{a_0,a_1,...,a_k} 
=\omega_{a_0 a} m^a_{a_1,...,a_k}$ is cyclically symmetric 
(in graded sense). The Chern-Simons functional  can be 
defined on ${\mathcal A}\otimes {\rm Mat }_N$ 
in the following way. In a basis $a_1,\dots,a_k$  for ${\mathcal A}$ the value of the functional on even $a=\sum a_it^i,t^i\in {\rm Mat }_N $ is equal to 
\[m(a)=\sum_k \pm m_{i_{1}\dots i_{k}}\tr (t^{i_1}\cdots  t^{i_k})\]
with  $\pm$ derived from the Koszul sign rules. 

We assume that the inner product is odd; then the Chern-Simons functional generates an odd  vector field $Q$ on $\Pi{\mathcal A}_N$ ; the conditions that should be imposed on the operations in $A_{\infty}$-algebra are equivalent to the condition $[Q,Q]=0$ or , in other words, to the condition that Chern-Simons functional obeys the BV master equation.

Notice that two quasi-isomorphic $A_{\infty}$-algebras 
are
physically equivalent (i.e. corresponding Chern-Simons
functionals lead to the same physical results).

A differential associative algebra can be considered as 
an $A_\infty$-algebra where only 
operatious $^{(1)} m$ and $^{(2)} m$ do not vanish; in 
this case both definitions of Chern-Simons functional 
coincide.

\section         {Observables of  Chern-Simons theory.}\label{S:Observables}

Recall that  in BV formalism  a physical theory is specified by an action functional $S$ defined on a space of fields  considered as  functions on odd symplectic manifold $\mathcal{E}.$ A classical observable is  defined as  an even  functional $a$ obeying $\{S,a\}=0$ or equivalently $\xi_Sa=0$ were $S$ is the action functional obeying the master equation and $\xi_S$ stands for the  operator $\xi_Sf=\{S,f\}$. (We restrict ourselves to the polynomial functionals.)  Classical 
observables 
are related to infinitesimal deformations of the solution to the master equation. In what follows we will consider also odd functionals $a$ obeying $\{S,a\}=0$ . These "odd observables" do not have a physical meaning of observables, but they correspond to odd infinitesimal deformations (deformations of the form $S+\epsilon a$ where $\epsilon$ is an odd parameter.)  Trivial observables (observables of the form $a=\{S,b\}$) correspond to trivial deformations (deformations induced by infinitesimal change of variables).
We will see that odd observables  generate even symmetry transformations and even observables generate odd symmetry transformations. 
  
Equivalently one can describe a theory by means  of  a homological  vector field $Q$ on $\mathcal{E}$  preserving the odd symplectic form; then the first order differential operator $\hat Q$ corresponding to $Q$ can be represented in the form $\xi_S$. Notice that the set of of observables depends only on $Q$ (the symplectic form and the action functional are irrelevant). The field $Q$  determines the equations of motion (a solution of equation of motion can be interpreted as a point in the zero locus of $Q$.)  If there exists a $Q$-invariant symplectic form these equations of motion come from action functional.

If  Chern-Simons theory is constructed by means of 
associative graded differential algebra ${\mathcal A}$ 
with an odd inner product   then  every   element of  cyclic
cohomology of ${\mathcal A}$ specifies an observable \cite{Sch}.  This fact follows from  the  
statement that infinitesimal deformations 
of ${\mathcal A}$ into $A_{\infty}$-algebra with  inner 
product are labelled by cyclic cohomology $HC({\mathcal A})$ 
of ${\mathcal A}$ \cite {Penkava}.  Algebra ${\mathcal A}$ determines Chern-Simons 
theory  for all $N$; the 
observables we are talking about  are defined  for 
every $N$.

The observables we consider can be constructed directly, without reference to \cite{Penkava}.
By definition a cyclic cocycle on ${\mathcal A}$ is a polylinear map $\sigma :{\mathcal A}^n\to {\mathbb{C}}$ satisfying some conditions.  Such a map can be used to construct a polylinear map $\sigma_N:  \mathcal{A}_N^n\to \mathbb{C} $ specifying a cyclic cocycle of $\mathcal{A}_N$ .This follows from Morita invariance of cyclic cohomology( \cite{LodayQuillen}). It is easy to give an explicit formula for $\sigma_N$, namely
$$\sigma _N(A_1,...,A_n)=\sum _{i_1,...,i_n} \sigma (a^{i_1,i_2}_1,a^{i_2,i_3}_2...,a^{i_n,i_1}_n), A_k=(a^{i,j}_k)$$
 An observable corresponding to the cyclic cocycle is the functional $\sigma _N(A,...,A)$ defined on $\Pi \mathcal{A}_N.$

Obviously a product of observables is an observable, hence elements of  $\Sym \Pi HC({\mathcal A})=\sum \Sym ^k\Pi HC({\mathcal A})$ specify observables defined for every $N$. Let us prove that all observables defined for every $N$ are of this kind. To classify observables we should compute the cohomology of  the first order differential operator ${\hat Q}=\xi_{CS}$ induced by the vector field  $Q$ in the space of polynomial functionals on  ${\mathcal A}_N$.   It is easy to check that up to parity reversion this operator can be identified with the differential  in the definition of Lie algebra cohomology of $L_N\mathcal{A}=Lie {\mathcal A}_N.$ It is well known \cite{LodayQuillen} that  for large $N$ the  cohomology of ${\mathcal A}$ with trivial differential is isomorphic to 
 $\Sym[\Pi  HC({\mathcal A})].$ 

 For any dga $\mathcal{A}$ there exists a homomorphism of $\Sym [\Pi  HC({\mathcal A})]$ into the projective (inverse) limit of groups $H(L_N\mathcal{A})$; in the cases we are interested in this homomorphism is an isomorphism (we do not know whether this is true in general).
 
\section {Ten-dimensional SUSY YM theory as generalized Chern-Simons  theory.}

We have constructed Chern-Simons theory starting with differential associative algebra $\A$ equipped with closed trace $\rm {tr}$ that generates  invariant  inner product   $<a,b>={\rm tr} ab$.   We have assumed that the inner product is non-degenerate; however,  one can show  that it suffices to assume the non-degeneracy of the induced inner product on homology.  This remark allows us to consider ten- dimensional SUSY YM theory and its dimensional reductions as generalized Chern-Simons theory.  

We define Berkovits algebra $B$  as the algebra of polynomial functions of pure spinor $
\lambda$, odd spinor $\psi$ and $x=(x^1,...,x^{10})\in \mathbb{R}^{10}$. 
Sometimes it is convenient to modify this definition considering an algebra $B^{\infty}$ consisting of functions that are polynomial in  $
\lambda$ and $\psi$ but smooth as  functions of  $x\in \mathbb{R}^{10}$. 
The differential is 
defined as the derivation 
\begin{equation}\label{E:differential}
d=\lambda^{\alpha}\bigg(\pr{\psi^{\alpha}}+\Gamma_{\alpha\beta}^i\psi^{\beta}\pr{x^{i}}\bigg).
\end{equation}

The algebra $B_d$ (Berkovits algebra reduced to $d$-dimensional space) is the algebra of functions depending on pure spinor $
\lambda$, odd spinor $\psi$ and $x=(x^1,...,x^{d})\in \mathbb{R}^{d}$. 
The differential is defined by the same formula.

One can use (\ref {Q}) to define an odd vector field $Q$ on $B_d\otimes Mat_N. $ It is well known  (\cite{AFG}, \cite{Howe},\cite{Berkovits}) that for every solution $A_{\alpha} (x,\theta)$ of equations of motion of ten-dimensional SUSY YM one can construct a point $\lambda ^{\alpha}A_{\alpha} (x,\theta)\in B_{10}\otimes Mat_N$ belonging to the zero locus of $Q$. ( We are working in superspace formalism, the superfield $A_{\alpha}$ takes values in $N\times N$ matrices.)  Starting with a solution of equations of motion  reduced to $d$-dimensional space  we obtain a point of the zero locus of the 
vector field $Q$ on $B_d\otimes Mat_N. $ 

For $d=0$ we obtain the reduced Berkovits algebra related to ten-dimensional SUSY YM theory reduced to a point. Introducing an appropriate trace we can apply the techniques of generalized Chern-Simons theory to this algebra. To apply these techniques to the case $d>0$ we should introduce the notions of local observables and local trace (see the next section).

\section{Lie algebra of local observables in the classical BV formalism}\label{S:local}

 Let   $f:M\rightarrow N$ be a $\mathbb{C}$-linear map of graded $R$-modules where $R$ is a graded commutative algebra over $\mathbb{C}$. We define $[x,f]:M\rightarrow N$ by the formula $xf(m)-(-1)^{|f||x|}f(xm),x\in R$. The map $f$ is local if $[x_1,\dots,[x_n,f]\dots]\equiv 0$ for some $n$ and all $x_1\dots,x_n\in R$ . If $R$ is an algebra of smooth functions on $\mathbb{R}^{n|k}$  and $M, N$ are the space of sections of finite rank vector bundles then  $f$ defines a  differential operator between vector bundles.
 
  If  $R,M,N$ are differential graded objects then the definition can be weakened  by replacing $[x_1,\dots,[x_n,f]\dots]\equiv 0$ by $[x_1,\dots,[x_n,f]\dots]\equiv [d,g]$ for some $g:M\rightarrow N$.
 
  Let $A$ be an associative graded algebra over $R$ and $M$ is a graded $A$-bimodule. The graded space of Hochschild cochains $C^k(A,M)=\Hom_{\mathbb{C}}(A^{\otimes k},M)$ contains a subspace $C_{R_{loc}}^k(A,M)$ of cochains $\cO$ that are local with respect to all variables (i.e.,   $f(x)=\cO(a_1,\dots,a_{i-1},{x},a_{i+1},\dots,a_k)$ is a local map of $R$-bimodules for $i=1,\dots,k$). 
 We omit the standard definition of the Hochschild differential referring to \cite{Abbaspour} or to \cite{Loday}.  By definition multiplication and the differentials in $A$ and $M$ are local. This is why $\prod_k C_{R_{loc}}^k(A,M)$ is a subcomplex in $\prod_k C^k(A,M)$. The cohomology of $\prod_k C_{R_{loc}}^k(A,M)$ is denoted by $\rHH^{k}_{R_{loc}}(A,M)$.  When no confusion is possible we will drop the $R$-dependence in the cohomology: $\rHH^{k}_{{loc}}(A,M)=\rHH^{k}_{R_{loc}}(A,M)$
 
 When $A=R$ is an  algebra of functions on  a smooth or an affine algebraic manifold $X$ 
and $M$ is space of sections of a vector bundle over $X$
then $\rHH^{k}_{loc}(A,M)=\rHH^{k}(A,M)$ (c.f. \cite{KontsDefo},Section 4.6.1.1.)

In the following we will assume that $R$ is an algebra of functions on  a smooth or an algebraic supermanifold $X$. This allows us to define the complex of integral forms $\{\Omega_{R}^{-i}\}$, equipped with the de Rham differential $d_{dr}$.

  By definition a local trace is  a series of graded local  maps $\tr_{i}: A\rightarrow \Omega_R^{-i}$ that for $i\geq 0$  satisfy 
 \[\tr_{i}(d_A a)=-d_{dr}\tr_{i+1}( a), \tr_{i}([a,b])=0.\]
When $R=\mathbb{C}$ this becomes a definition of an ordinary trace that determines an inner product $tr(ab)$ on $A$. This inner product  allows us to define a map of cohomology groups $\rHH^i(A,A)\rightarrow \rHH^{i}(A,A^*)$. Our next set of definitions is intended to create a setting which would accommodate a similar homomorphism in a local setting.

 The complex $C^k(A,A^*)$ coincides with the space of graded maps $\Hom(A^{\otimes (k+1)},\mathbb{C})$. The complex is equipped with the  differential $d_{hoch}$  (see \cite{Abbaspour}, \cite{Loday}  for details) that computes Hochschild cohomology $\rHH^{i}(A,A^*)$.  
  By definition the complex $C_{loc}^k(A)$ is a subcomplex of the complex $\prod_{ij}\Hom_{\mathbb{C}}(A^{\otimes {k+1}},\Omega_R^{-i})$ of  $R$- local maps. The differential $d$ is equal to the sum $d_{hoch}+d_{dr}$, where $d_{dr}$ is acting in $\bigoplus_i \Omega_R^{-i}$. We denote the cohomology of $C_{loc}^k(A)$ by $\rHH_{loc}^{k}(A)$. Note that when $R=\mathbb{C}$ the cohomology  $\rHH_{loc}^{k}(A)$ becomes equal to  $\rHH^{k}(A,A^*)$
 
 The $R$-local cyclic cohomology are defined along the same line: the subspace  $\prod_k CC^k(A)$  of  $C_{loc}^k(A)$ consisting of cyclic cochains is a subcomplex. Corresponding cohomology is denoted by  $\rHC_{loc}^{n}(A)$.

 The groups $\rHH_{loc}^{n}(A,A)$, $\rHH_{loc}^{n}(A)$  and $\rHC_{loc}^{n}(A)$ have "multi-trace" generalizations.
 By definition the chains of $\rHH_{loc,mt}^{n}(A,A)$ is a subspace of the $R$-local maps in $\Hom_{\mathbb{C}}(T(A)\otimes \Sym[\Pi CC(A)],A)$. The space of chains for $\rHH_{loc,mt}^{n}(A)$ is a subspace of local maps in $\Hom_{\mathbb{C}}(T(A)\otimes A\otimes \Sym[\Pi CC(A)],\Omega_R)$, the space  of chains for $\rHC_{loc,mt}^{n}(A)$ is a subspace of local maps in $\Hom_{\mathbb{C}}( \Sym[\Pi CC(A)],\Omega_R)$. Keep in mind that such generalizations  for the classical nonlocal theories reduce to the  space of linear maps of  $\bigoplus_k\rHH^{k}(A,A)$ and  $\bigoplus_k\rHH^k(A,A^*)$ with  $\Sym[\Pi\rHC(A)]$.
 
 The local version of the cyclic and  Hochschild cohomology shares many common properties with their classical counterparts. In particular there is a long exact sequence
 
 \[\cdots \rightarrow \rHH_{loc}^{n}(A) \rightarrow  \rHC_{loc}^{n-1}(A) \rightarrow \rHC_{loc}^{n+1}(A) \rightarrow \rHH_{loc}^{n+1}(A) \rightarrow  \cdots \]
 There is a similar sequence for the multi-trace version of the theories.
 
 A  local trace functional $\{\tr_I\}$, which we defined above,  is  a cocycle in $\rHC_{loc}^0(A)$. It defines  maps $\rHH_{loc}^{s}(A,A)\rightarrow \rHH_{loc}^{s}(A),\rHH_{loc,mt}^{s}(A,A)\rightarrow \rHH_{loc,mt}^{s}(A)$ by the formula
 \[c(a_1,\dots,a_s)\rightarrow \sum_{i}\tr_{i}(c(a_1,\dots,a_s)a_{s+1}).\]
  We say that the trace $\tr=\{\tr_i\}$ is homologically nondegenerate if it induces  an isomorphism $\rHH_{loc}^{s}(A,A) \rightarrow \rHH_{loc}^{s}(A)$.

Some of the  discussion from  Section \ref{S:Observables} can be carried out in the local setting. In particular local vector fields, local functionals and local k-forms  obviously make sense. There are homomorphisms
\[\rHH_{loc,mt}^{n}(A,A)\rightarrow \rH^n(L_NA,L_NA),\] \[\rHC_{loc,mt}^{n}(A)\rightarrow \rH^n(L_NA),\] \[\rHH_{loc,mt}^{n}(A)\rightarrow \rH^n(L_NA,L_NA^*) \]

The relative Hochschild cohomology group $\rHH_{R}^{n}(A,A)$ maps to $\rHH_{loc}^{n}(A,A)$. The relations of the relative groups $\rHH_{R}^{n}(A,A^*)$  and $\rHC_{R}^{n}(A)$ to $\rHH_{loc}^{n}(A)$  and $\rHC_{loc}^{n}(A)$ is obscure.

The last type of generalization is intended for a dga $A$ without a unit  over a commutative unital dga $R$. The definitions follow closely the above outline.  The reader can consult on  the details  of cohomology theory  of algebras without a unit in \cite{Loday}.

\section {Symmetry preserving deformations}

As we mentioned already  infinitesimal  deformations  of the solution of master equation correspond to observables. Observables belonging to the same cohomology class specify equivalent deformations, i.e. deformations related by a change of variable (by a field redefinition).
This means that in BV formalism the deformations of physical theory with action functional $S$ are labeled by the cohomology of the differential $\xi_S$.

As we have seen an even element of $\Lambda  HC(\mathcal{A})$ specifies an observable of Chern-Simons theory defined for every $N$; hence it determines an infinitesimal  deformation  of Chern-Simons theory defined for all $N.$

We will be interested in the deformations of Chern-Simons theory that  are defined for every $N$ and preserve the symmetry of original theory.

Let us make some general remarks about symmetries in BV formalism. If the equations of motion are specified by a homological vector field $Q$ then every vector field $q$ commuting with $Q$ determines a symmetry of equations of motion. (The vector field $q$ is tangent to the zero locus of $Q$.)  The vector field $q$ can be even or odd; in other words we can talk about super Lie algebra of symmetries. However, among these symmetries there are trivial symmetries, specified by vector fields of the form $[Q,a]$ where $a$ is an arbitrary vector field. This means that the super Lie algebra $\mathcal{L}$ of non-trivial symmetries of EM can be described as homology of the space of all vector fields with respect to the differential defined as a commutator with $Q$. If we would like to consider only Lagrangian symmetries, i.e. symmetries corresponding to vector fields $q$ having the form $\xi _s$
we obtain a Lie algebra $\widetilde {\mathcal{L}}$ isomorphic to the homology of operator $\hat Q$. %In other words, the elements of $\{S,a\}=0$ 
Notice, that both $\mathcal{L}$ and  $\widetilde {\mathcal{L}}$ depend only on the vector field $Q$. However,  the natural homomorphism  $\widetilde {\mathcal{L}}\to \mathcal{L}$ does depend on the choice of odd symplectic structure on $\mathcal{E}.$

Let us calculate $\mathcal{L}$ and  $\widetilde {\mathcal{L}}$ in the case of Chern-Simons theory restricting ourselves to the symmetry transformations, that can be applied for all $N$. The calculation of $\widetilde {\mathcal{L}}$ coincides with calculation of observables (up to parity reversal); we obtain
$\widetilde {\mathcal{L}}=\Lambda HC({\mathcal A}).$

To calculate $\mathcal{L}$ we  should study the cohomology $H(L_N(\mathcal {A}), L_N(\mathcal{A}))$ (the cohomology of the Lie algebra $L_N(\mathcal{A})$ with the coefficients in adjoint representation).  In the case of algebra $\mathcal{A}$ with zero differential these cohomology are well known  for large $N$ (\cite{Goodwillie}, \cite{LodayQuillen}) ; they are equal to
$$\Sym[\Pi  \rHC({\mathcal A})]\otimes \rHH(\mathcal{A},\mathcal{A}) .$$

 For dga algebras we  have maps 
 \[ \rH(L_N\mathcal{A},L_N\mathcal{A})\leftarrow \Sym[\Pi  \rHC({\mathcal A})]\otimes \rHH(\mathcal{A},\mathcal{A})\]
In interesting cases the LHS stabilizes for large $N$ and becomes isomorphic to the RHS. It is not clear whether this situation is general. 

We can say that $\g$ is a symmetry Lie algebra if it is embedded into $\mathcal{L}$ (or into $\widetilde {\mathcal{L}}$ if we would like to consider only Lagrangian symmetries). Fixing some system of generators $e_{\alpha}$ in $\g$  with structure constants $f_{\alpha \beta}^{\gamma}$ one  can  say that $\g$ is a symmetry Lie algebra of BV theory with homological vector field $Q$ if there exist  symmetry transformations $q_{\alpha}$ satisfying commutation relations 
\begin{equation}
\label{q}
[q_{\alpha},q_{\beta}]=f_{\alpha \beta}^{\gamma}q_{\gamma}+[Q,q_{\alpha \beta}]
\end{equation}for some vector fields $q_{\alpha \beta}$. However, it is useful  to accept a more restrictive definition of symmetry Lie algebra. We will say that $\g$ is a symmetry algebra if we have
an $L_{\infty}$-homomorphism of $\g$ into the differential Lie algebra of vector fields  (this Lie algebra is equipped with a differential defined as a commutator with $Q$). 
$L_{\infty}$-homomorphism of Lie algebra $\g$ with generators $e_{\alpha}$ into differential Lie algebra $\mathcal{V}$ is defined as a sequence $q_{\alpha},q_{\alpha_1,\alpha_2},....\in \mathcal{V}$ obeying some relations, generalizing (\ref {q}). (See \cite{LadaStasheff} for details.) 

Let us suppose that Lie algebra $\g$ acts on the differential algebra $\mathcal{A}.$ This means that we have fixed a homomorphism $\phi:\g\to Der \mathcal{A}$ of $\g$ into the Lie algebra of derivations of $\mathcal{A}$.  (It is sufficient to assume that we have an $L_{\infty}$action, i.e. an $L_{\infty}$ homomorphism of $\g$ into differential algebra  $Der(\mathcal{A}).$) This action  specifies $\g$ as a Lie algebra of symmetries of Chern-Simons functional    for every $N$. We are interested in infinitesimal deformations of this functional preserving these symmetries ($\g$-invariant deformations).  We identify two deformations related by the change of variables.

The space $C^{\bullet}(\mathcal{A},\mathcal{A})$ of Hochschild cochains  with coefficients in $\mathcal{A}$ has a natural $L_{\infty}$ action of $\g$, hence we can consider the cohomology of Lie algebra $\g$ with coefficients in this module. We will denote this cohomology by $HH^{\bullet}_{\g}(\A,\A)$ and call it Lie -Hochschild cohomology with coefficients in $\A .$ (For trivial $\g$ it coincides with Hochschild  cohomology of $\A$,  for trivial $\A$ with Lie algebra cohomology of $\g.$) Similarly we can define $HC^{\bullet}_{\g}(\A)$ (Lie-cyclic cohomology), $HH_{\g}^{\bullet}(\A,\A ^*)$ . There are also multi-trace version of these groups. For example the multi-trace version  of $HC^{\bullet}_{\g}(\A)$ uses uses the symmetric algebra of standard cyclic bicomplex $CC^{i}(\A)$. The multi-trace cyclic  cohomology group $HC^{\bullet}_{\g,mt}(\A)$ is the cohomology of the bicomplex $C^{\bullet}(\g,\Sym[CC^{\bullet}(\A)])$. The multi-trace equivariant version of Hochschild cohomology is cohomology of the 
tri-complex $C^{\bullet}(\g,\Sym[CC^{\bullet}(\A)]\otimes C^{\bullet}(\A,\A))$
One can prove the following theorem:

The $\g$-invariant deformations of Chern-Simons action functional $CS(\A)$ that are defined for all $N$ simultaneously  are labelled by the elements of $\rHC^{\bullet}_{\g,mt}(\A)$.

We can use this theorem to study supersymmetric deformations of ten-dimensional SUSY YM theory represented as Chern-Simons theory corresponding to the Berkovits algebra.

There is a number of modifications of the Berkovits algebra $B_{10}$ that depend on smoothness of its elements as  functions on $\mathbb{R}^{10}$ and their asymptotics at infinity.  Possible choices are polynomials functions , which are elements of  $\mathbb{C}[x_1,\dots,x_{10}]$.  This way we get $B^{poly}_{10}$. Similarly we can get an analytic modification $B^{an}_{10}$ which contains the  algebra of analytic functions $C^{an}(\mathbb{R}^{10})$ or the smooth version $B^{\infty}_{10}\supset C^{\infty}(\mathbb{R}^{10})$ , with or without restriction on asymptotics at infinity.  
Our following computations don't depend on what pair of algebras 
\[\begin{split}
&R=\mathbb{C}[x_1,\dots,x_{10}]\subset B^{poly}_{10}=A \\
&R=C^{an}(\mathbb{R}^{10})\subset B^{an}_{10}=A \text{ or }\\
&R=C^{\infty}(\mathbb{R}^{10})\subset B^{\infty}_{10}=A.
\end{split}
\]
we choose for cohomology computations. This is why we will use $B_{10}$ as a unifying notation for all  modifications.
 
 We can calculate  groups $\rHH^{i}_{loc\ \susy}(B_{10},B_{10})$. 
These linear spaces have an additional conformal grading  by eigenvalues of the dilation operator scaled by the factor of two\footnote{We do this to avoid fractional gradation in spinor components.}:  $\rHH^{i}_{loc\ \susy}(B_{10},B_{10})= \bigoplus_{k\in \mathbb{Z}} \rHH^{i,k}_{loc\ \susy}(B_{10},B_{10})$.

They can be expressed in terms of the groups $\rH^{s,t}(L,U(TYM))$ considered in \cite{M4}:
$$\rHH^{i,k}_{loc\ \susy}(B_{10},B_{10})=\rH^{i+k,i}(L,U(TYM))$$

The groups $H^{k,t}(L,U(TYM))$ were  calculated in \cite{M4} for $k=2.$ (See also \cite {MSSUSY}.) Similar methods can be applied for other values of $k.$

\section {Construction of deformations}
One can construct some interesting symmetry preserving deformations  starting with homology classes of symmetry Lie algebra $\g.$

 The application   of the homology of $\g$ to the analysis of deformations is based on the construction of the homomorphism
 \begin{equation}
\label {psi}
\psi:H_i(\g, N)\to H^{s-i} (\g, N)
\end{equation}
 for arbitrary differential module $N$ with  L$\ity$ action of $\g.$   Here $s$ stands for the number of even generators of $\g.$ This  homomorphism  is described in the appendix.
 
We will apply the homomorphism $\psi$ to the construction of symmetry preserving deformations.

Let $\A$ be a differential  $\mathbb{Z}$-graded associative algebra. {\footnote { One can generalize our constructions to the case of 
A$\ity$ algebras using the fact that a $\mathbb{Z}$-graded A$\ity$ algebra is quasi-isomorphic to differential graded algebra,}}
 Let us assume that $A$ is equipped with L$\ity$ action of Lie algebra $\g$ . Then $C^{\bullet}(A,A^*)$ and $HH^{\bullet}(A,A^*)$ are   L$\ity$ $\g$- modules and we can talk about the homology and cohomology of $\g$ with coefficients in these modules. (Recall that $C^{k}(A,A^*)$ stands for the module of Hochschild co-chains, i.e. of $k$-linear functionals on $A$ with values in $A^*$. Notice, that these co-chains can be identified with $(k+1)$-linear functionals with values in $\mathbb{C}.$) 
The complex $C^{\bullet}(A,A^*)$ has an additional operation of degree minus one :\[C^{\bullet}(A,A^*)\rightarrow C^{\bullet-1}(A,A^*)\]-the Connes differential $B$ .
The map $B$ is a composition of two maps $\alpha B_0$, which   look particularly simple if the degrees of all elements are even. The operator $\alpha$ is the operator of cyclic antisymmetrisation. The operator $B_0$ is defined by the formula \[(B_0\psi)(a^0,\dots,a^n)=\psi(1,a^0,\dots,a^n)-(-1)^{n+1}\psi(a^0,\dots,a^n,1)\] The reader may  consult \cite{Connes}, \cite{Loday} for details. {\footnote {The operator $B$
transforms Hochschild $n$-cochain into cyclic $(n-1)$-cochain; it generates a homomorphism
$HH^n(A,A^*)\to HC^{n-1}(A)$ in Connes exact sequence.}}

This operator  induces  map on $C_{\g}^{i}(A,A^*)$ and $C^{\g i}(A,A^*)$, denoted by the same symbol; it   anticommutes with $d_{\g}$ and $d_c$. 

The Connes operator induces a differential on Hochschild cohomology.
Let us assume in that the cohomology of $B$ in $H^{i}(A,A^*)$ is  trivial for $i>0$ and is one- dimensional for $i=0$. \footnote { This is true , for example, if there exists an auxiliary grading by means of non-negative integers with one-dimensional grading zero component. } This assumption permits us to construct an element
of  homology $H_{\bullet}(\g,C^{\bullet}(A,A^*))$
starting with any element $c_0\in H_{\bullet}(\g,C^{\bullet}(A,A^*))$ obeying $Bd_{\g}c_0=0$. 
The construction is based on the observation
that due to triviality of the cohomology of $B$ we 
can represent $d_{\g}c_0$ as $Bc_1$. Applying $d_{\g}$ to both parts of equation  $d_{\g}c_0=Bc_1$
we obtain $Bd_{\g}c_1=0$; this equality allows us to continue the process. The process will terminate when $d_{\g}c_i=0$.  This must happen for some $i$ because $d_{\g}$ decreases the degree in $\Sym (\Pi \g)$ (the number of ghosts).
The element $c_i$ specifies the homology class we are interested in.

Let us describe the construction of elements of $H_{\bullet}(\g,C^{\bullet}(A,A))$,  which uses homology classes of super Lie algebra $\g$ with trivial coefficients and a $\g$-equivariant trace $\tr$  as an input. (We assume that the trace specifies non-degenerate inner product on cohomology.) The trace determines a 
$\g$-equivariant map $A^*\to A$ and therefore
a homomorphism $H_{\bullet}(\g,C^{\bullet}(A,A^*))\to H_{\bullet}(\g,C^{\bullet}(A,A))$. This means that it is sufficient to construct an element of $H_{\bullet}(\g,C^{\bullet}(A,A^*))$.

Let us take  a  representative $c\in \Sym (\Pi \g)$ of homology class of the Lie algebra $\g$.
Then we can define $c_0$ by the formula  $c_0=c\otimes\epsilon$, where $\epsilon$ stands for 
a homomorphism of the algebra $A$ into a field. 
{\footnote { If $A$ is  represented as a direct sum of one-dimensional subalgebra generated by the unit and ideal $I$ (augmentation ideal) then $\epsilon $ is a projection of $A$ on the first summand.}} It is easy to check that $B\epsilon =0$ and that $\epsilon$ specifies a non-trivial  class in the homology of $B$.
We see that $Bd_{\g}c_0=-d_{\g}Bc_0=0$.  This means that we can apply the iterative construction described above to obtain a cycle $c_l$. The corresponding homology class
$[c_l]\in H_{\bullet}(\g,C^{\bullet}(A,A^*))$ is the class we need.

{\bf Remark}
Recall that we have defined $\g$-equivariant version of cyclic cohomology (Lie-cyclic cohomology) $HC^{\bullet}_{ \g}(\A)$ as cohomology of the Lie algebra $\g$ with coefficients in cyclic cochains considered as a differential $\g$-module. It is rather straightforward to construct Connes long exact sequence. The main corollary of this construction is that classes $c_l$ are images of classes in cyclic cohomology. 

{\bf Remark} The equivariant version of the package $\rHC^{\bullet}(\A)$, $\rHH^{\bullet}(\A,\A)$, $\rHH^{\bullet}(\A)$ makes sense for both of local versions described in Section \ref{S:local}.

\section {Appendix. Homology of super Lie algebras}
\subsection{Finite-dimensional super Lie algebras}
%\subsection{}

Let us consider first the cohomology of finite-dimensional super Lie algebras. This cohomology is defined in terms of a differential 
$$d=\frac{1}{2}(-1)^{|b^k|}f_{lk}^mb^kb^lc_m$$
where $f_{kl}^m$ are structure constants of super Lie algebra $G$ in some basis $t_k$. The operators $b^k$ and  $c_k$ correspond to elements of basis, but have  parity  opposite to the parity of elements of basis. They satisfy canonical (anti)commutation relations:
$[c_k,b^l\}=\delta _k^l;$
in other words they can be considered as generators of super Weyl algebra $W_{rs}$ where $r$ stands for the number of even generators and $s$ stands for the number of odd generators .The differential acts in   any representation of Weyl algebra (in any $W_{rs}$-module);  the cohomology can be defined by means of any representation  and depends on the choice of representation. We will assume that the representation $\mathcal F$ of Weyl algebra $W_{rs}$ is graded  in such a way that $b^k$ raises grading by $1$ and $c_k$ decreases grading by $1$, then the differential increases grading by $1$. The cohomology is also graded in this case. One can define  cohomology of super Lie algebra $G$ with coefficients in $G$-module $N$ by means of the differential $d+\sum T_kb^k$ on the space $\mathcal F\otimes N$ (Here $T_k$  denotes the action of generator $t_k\in G$ on $N$.) We will use the notation  $H^k(G,N|\mathcal F)$ for $k$-dimensional cohomology of super Lie algebra $G$ with coefficients in $G$-module $N$ calculated by means of $W_{rs}$-module $\mathcal F$.

 If $G$ is a conventional Lie algebra then $r=0$, the super Weyl algebra is a Clifford algebra with $2s$ generators. Irreducible representation in this case is unique; the representation space can be realized as Grassmann  algebra (as algebra of functions of $s$ anticommuting variables) where $c_k$ and $b^k$ act as derivatives and multiplication operators. We come to the standard notion of cohomology of Lie algebra.(However, in the case when $G$ is an infinite-dimensional Lie algebra the irreducible representation of corresponding Clifford algebra is not unique; this remark leads to the notion of semi-infinite cohomology.) If $r>0$ the super Weyl algebra is a tensor product of Weyl algebra $W_r$ and Clifford algebra $Cl_s$; further $W_r$ is a tensor product of $r$ copies of Weyl algebra $W=W_1.$ {\footnote {Notice that we work in algebraic setting. It is well known that for correct definition of unitary representation Weyl algebra with finite number of generators has only one irreducible unitary representation. This statement cannot be applied to representations at hand.}}

Let us consider first of all representations of the algebra $W$ having generators $b,c$ with relation $[c,b]=1.$ The simplest of these representations $F_+$ is realized in the space of polynomials $\mathbb {C}[t]$ where $c$ acts as a derivation and $b$ as a multiplication by $t$. The grading is given by the degree of polynomial. This representation can be described also as representation with cyclic vector $\Phi$ obeying $c\Phi=0$ (Fock representation with vacuum vector $\Phi.$) Another representation $F_-$ can be constructed as a representation with cyclic vector $\Psi$ obeying
$b\Psi=0$, $\deg \Psi=0$ . To relate these two representations we consider the representation $F$ in the space $\mathbb {C}[t,t^{-1}]$ ( polynomials of $t$ and $t^{-1}$). The operators $c$ and $b$ again act as derivation and multiplication by $t$. It is easy to check that factorizing $F$ with respect to subrepresentation $F_+$ we obtain a representation isomorphic to $F_-$ (the polynomial $t^{-1}$ plays the role of cyclic vector $\Psi$). Notice, however, that the grading in $F_-$ does not coincide with the grading in $F/F_+$ (the degree of $t^{-1}$ is equal to $-1$). One can say that as graded module $F/F_+$  is isomorphic to $F_-[-1]$ (to $F_-$ with shifted grading).

Let us represent $W_{rs}$ as a tensor product $W\otimes W_{r-1,s}$.For every representation $E$ of second factor we can construct two representations of $W_{rs}$ as tensor products
$F_+\otimes E$  and $F_-\otimes E$.The relation $F/F_+=F_-[-1]$ permits us to construct a map 
\begin {equation}
\label {p}
H^k(G,N|F_-\otimes E)\to H^{k}(G, N|F_+\otimes E)
\end{equation}
This map is analogous to picture changing operator in BRST cohomology of superstring. It can be regarded as coboundary operator in exact cohomology sequence corresponding to short exact sequence
$$0\to F_+\otimes E\to F\otimes E\to F_-\otimes E\to  0.$$
Notice that coboundary operator raises degree by $1$, but taking into account the the shift of grading in $F_-$ we see that ( \ref {p}) does not change the degree.

We will consider irreducible representations $F_{\epsilon _1, ..., \epsilon _r}$ of $W_{rs}$ defined as tensor product of representations $F_{\epsilon _k}$ and irreducible representation of Clifford algebra. (Here $\epsilon _k=\pm$.)These representations can be defined also as Fock representations with vacuum vector $\Phi$ obeying $c_k\Phi=0$ if $\epsilon _k=+$ and $b^k\Phi=0$ if $\epsilon _k=0.$ The grading  is determined by the condition $\deg \Phi=0.$ The cohomology corresponding to representation 
with all $\epsilon _k=+$ coincides with standard cohomology of super Lie algebra:
$$H^k(G,N|F_{+...+})=H^k(G,N).$$ 
The cohomology corresponding to representation with all $\epsilon _k=-$ is closely related to homology of super Lie algebra. If $f_{kl}^l=0$ we have
\begin{equation}
\label{h}
H^k(G,N|F_{-...-})= H_{s-k}(G,N).
\end{equation}
To check  ( \ref {h}) we notice that  homology can be defined by means of differential
$$\partial=\frac{1}{2}(-1)^{|\gamma_k|}f_{lk}^m\gamma _m\frac{\partial ^2}{\partial \gamma_ k \partial\gamma _l}+ T_k \frac{\partial}{\partial \gamma _k}$$
acting in the space of polynomial functions of variables $\gamma _k$ (ghost variables). (Here as earlier $T_k$ denotes the action of $t_k\in G$ on $G$-module $N$. The ghost variables have the parity opposite to the parity of $t_k$.)  We can rewrite this differential in the form 
$$\partial =\frac{1}{2}(-1)^{|b_k|} f_{lk}^mc_mb^kb^l+T_kb^k$$
where $c_k,b^l$ satisfy canonical (anti)commutation relations. If $f_{kl}^l=0$  the differential takes the form of cohomology differential acting in the space $F_{-...-}$. (The constant polynomial is a cyclic vector $\Phi$ obeying $b_k\Phi =0$.) However, the grading is different: in the space of polynomial functions of $\gamma _k$ the operator $c_k$ increases degree by $1$ (instead of decreasing it by $1$ in cohomological grading). The grading of the cyclic vector $\Phi$ (of the Fock vacuum) is also different in homological and cohomological setting ($0$ versus $s$).  We obtain the formula (\ref {h}).

Applying $r$ times the homomorphism (\ref {p}) we obtain a homomorphism from homology into cohomology. More precisely, if $f_{kl}^l=0$ we obtain a homomorphism
$$ H_i(G,N)\to H^{s-i}(G,N)
$$.
\subsection {Infinite-dimensional super Lie algebras}
Recall that the cohomology of finite-dimensional super Lie algebra $G$ with coefficients in $G$-module $N$ were defined by means of differential 
\begin{equation}
\label{d}
d=\frac{1}{2}(-1)^{|b_k|} f_{lk}^mb^kb^lc_m+T_kb^k
\end{equation}

acting on the tensor product  $\mathcal{F}\otimes N$. Here $\mathcal{F}$ is a representation of super Weyl algebra with generators $b^k,c_l$ , the symbol $f_{kl}^m$ denotes structure constants of $G$ in the basis $t_k$ and $T_k$ stands for the operator in $N$ corresponding to $t_k$,
The space $\mathcal{F}$ can be considered as a $G$-module; the elements $t_k\in G$ act as operators
\begin{equation}
\label{t}
\tau _k=f_{kl}^mb^lc_m.
\end{equation}

These operators obey  relations
\begin{equation}
\label{tau}
[\tau _k,\tau _l\}=f_{kl}^m\tau _m,
\end{equation}
\begin{equation}
\label{tb}
[\tau _k,b^m\}=f_{kl}^mb^l,
\end{equation}
\begin{equation}
\label{tc}
[\tau _k, c_l\}=f_{kl}^mc_m.
\end{equation}
The differential $d$ obeys
\begin{equation}
\label{db}
[d,b^m\}=\frac{1}{2}f_{kl}^mb^kb^l,
\end{equation}
\begin{equation}
\label{dc}
[d,c_l\}=\tau _l+T_l.
\end{equation}
Let us consider now the case when $G$ is an infinite-dimensional super Lie algebra and $N$ is a projective representation of $G$ (=a module over central extension of $G$). We will keep the notation $t_k$ for the elements of basis of $G$ and $f_{kl}^m$ for structure  constants. {\it We will assume that for fixed indices $k,l$ there exists only finite number of indices $m$ such that $f_{kl}^m\neq 0.$ Similarly, if indices $k,m$ are fixed then $f_{kl}^m\neq 0$ only for finite number of indices $l.$}
The formulas  (\ref {d}) and (\ref {t}) in general do not make sense in this situation. However, the RHS
of (\ref {tb}) and (\ref {tc}) is well defined. We will assume $\mathcal {F}$ is an irreducible representation of Weyl algebra; then these formulas specify $\tau _k$ uniquely up to an additive constant. If the solution for $\tau _k$ does exist it specifies a projective representation of $G$:
$$[\tau _k,\tau _l\}=f_{kl}^m\tau _m+\gamma _{kl}.$$  The constants $\gamma _{kl}$ determine a two-dimensional cocycle of $G$; in physics it is related to central charge. 

{\it We assume that  the two-dimensional cohomology class of $G$ corresponding to the projective module $N$ is opposite to the cohomology class of $\gamma$. This means that for appropriate choice $\tau _l$ the expression  $\tau _l+T_l$ (the RHS of (\ref {dc})) specifies a genuine representation of $G$}

We will consider the case when $\mathcal{F}=\mathbb{F}_I$ is a Fock module ( a module with a cyclic vector $\Phi$ obeying $b^k\Phi=0$ for $k\in I$, $c_l\Phi=0$ if $l\in J$ where $J$ denotes the complement to $I$). Here $I$ stands for some set of indices;{\it we assume that there exists only a finite set of triples $(k,l,m)$ with $f_{kl}^m\neq 0$ obeying $k,l\in J, m\in I.$} {\footnote {In many interesting situations $G$ as a vector space can be represented as a direct sum  of two subalgebras; the representation of the set of indices as a a disjoint union of $I$ and $J$ is related to this decomposition. In this case the cohomology we are interested in is called semi-infinite cohomology.}}Then  $\tau _k$ obeying equations (\ref {tb}) and (\ref {tc}) can be written in terms of normal product
\begin{equation}
\label{tn}
\tau _k=f_{kl}^m:b^lc_m:    .
\end{equation}
Under our assumptions the RHS of (\ref {db}) and (\ref {dc}) specifies a well defined operator on $\mathcal{F}.$  Considering these formulas as equations for $d$ we see that  they determine $d$ up to an additive  constant. Requiring $d^2=0$ we obtain the following expression for $d$:
\begin{equation}
\label{dd}
d=\frac{1}{2}(-1)^{|\gamma_k|} f_{lk}^m:b^kb^lc_m:+T_kb^k.
\end{equation}
One defines the cohomology of $G$  with coefficients in $N$ by means of differential $d$. The cohomology in general depends on the choice of set $I$ (on the choice of picture).  One can introduce grading in $\mathcal{F}$ assuming that $b^k$ increases degree by $1$ ,$c_l$ decreases degree by $1$ and that $\deg \Phi=0$. Using this grading (and grading in $N$) one can define grading on cohomology. We will use the notation $H^n(G,N;I)$ for $k$-dimensional cohomology.

We would like to study relation between $H^n(G, N;I)$ and $H^n(G,N;I')$ (the dependence of cohomology on the choice of the picture). We will analyze the case when $I'$ is obtained from $I$ by deleting one index $k$. Let us notice first of all that in the case when  $t_k$ is an even generator
(corresponding ghosts are odd)  $H^n(G, N;I)=H^n(G,N;I')$; this follows from  the fact that $\mathbb{F}_I$  can be identified with $\mathcal{F}'.$ (If a vector $\Phi$ obeys $b^k\Phi=0$ for $k\in I$, $c_l\Phi=0$ for $l\in J$  the vector $\Phi'=c_k\Phi \in \mathcal{F}_I$ obeys $b^k\Phi'=0$ for $k\in I'$, $c_l\Phi=0$ if $l\in J'$ where $J'$ stands for the complement of $I'.$) If the generator $t_k$ is odd (corresponding ghosts are even) then repeating the arguments used for finite-dimensional Lie algebras we can construct a homomorphism
 \begin{equation}
\label{pp}
H^n(G,N; I)\to H^n(G,n;I')
\end{equation}
(picture changing operator).

This homomorphism is not an isomorphism in general. However, it is an isomorphism in cases relevant for string theory (when $G$ is a superanalog of Virasoro algebra).
\vskip .5in
%\bibliographystyle{plain.bst}
%\bibliography{chernym}

\end{document}